\documentclass[twocolumn]{elsart}

\usepackage{natbib}
\usepackage{graphicx,color,psfig}

\usepackage{amssymb}

\newcommand{\ignore}[1]{}

\begin{document}

\begin{frontmatter}



\title{Parameters for Very Light Jets of cD Galaxies}


\author{M. Krause} \and
\author{M. Camenzind}

\address{Landessternwarte K\"onigstuhl, D-69117 Heidelberg, Germany}

\begin{abstract}
Recent Chandra X-ray observations of jets in central cluster galaxies show interesting 
features around the symmetry plane. We have carried out simulations 
involving bipolar jets, 
removing the artificial boundary condition at the symmetry plane. 
We use a very low jet density (IGM/jet $\approx 10^4$) and take into account a decreasing density 
profile. We find that the jet bow shock undergoes two phases: First a nearly spherical
one and second the well-known cigar-shaped one. We propose Cygnus A to be in a 
transition phase, clear signs from both phases. Due to inward growing
of Kelvin Helmholtz instabilities (KHI) between cocoon and shocked IGM, mass entrainment is 
observed predominantly in the symmetry plane. We propose this mechanism to
produce some of the so far 
enigmatic X-ray features in the symmetry plane in Cygnus~A. 
\end{abstract}

\begin{keyword}
Hydrodynamics \sep Simulation \sep Jet propagation
\PACS 95.30.L \sep 47.27.W \sep  95.30.Lz \sep 98.38.F \sep 98.54 \sep 98.58.F 
\sep 98.62.Nx \sep 98.62.Ra \sep 98.65.Cw \sep 98.70.Dk
\end{keyword}

\end{frontmatter}

\section{Introduction}
\label{intro}
Radio galaxies are traditionally known by their radio emission, which arises 
from magnetised jet plasma, accelerated to large velocities in the active nucleus,
and expelled up to Mpc distances. During their growth, they displace and compress
the ambient gas. In the central and dominant cluster galaxies (cDGs), 
this gas is usually dense enough to be seen in the X-ray regime
by bremsstrahlung.
The Chandra observatory can now resolve these centers of galaxy clusters
with unprecedented resolution. The displaced hot gas ($10^7$~K) has been detected 
in several cases \citep[e.g.][and references therein]{Sokea02}. 
An exceptional case is the Cygnus cluster
with its famous FR~II radio galaxy Cygnus~A. 
This is the only nearby (redshift z=0.06) powerful jet source.
In Cygnus~A, the leading bow shock can be found on the X-ray image 
\citep[][also compare Chandra homepage]{Sea01}. The identification is easy: 
It closely engulfs and follows the radio lobes in vicinity  of the hotspots,
and forms an almost spherical bubble around the radio emission.
It also reproduces the armlength asymmetry. From this data, the axis ratio
of the bow shock can be determined to 1.2. Comparison with the radio data
gives a bow-shock-to-cocoon width of 3.8.
In the symmetry plane, X-ray bright filaments are present, next to and 
within the radio lobe. Such excellent data demands detailed modelling.
We have simulated a jet with the conditions assumed to be present 
in Cygnus~A (sec.~\ref{3Dsim}). 
The simulation is bipolar 
in order to take into account interactions of the backflows of the jets.
Because it fails to reproduce
the data, quantitatively, we present a parameter study in sec.~\ref{pars}.
Finally, we present a modified jet model for Cygnus~A in sec.~\ref{discu}
and discuss the application to comparable jet sources.

\section{Bipolar 3D Simulation}
\label{3Dsim}
Using the code Nirvana \citep{ZY97}, we simulated \citep{myphd} a 
bipolar jet in 3D and cylindrical coordinates with the parameters of the Cygnus~A 
jet, estimated by \citet{CarBar96}: beam radius $r_{\mathrm{j}}=550$~pc,
velocity $v_\mathrm{j}=0.4c$, power $L_\mathrm{j}=10^{46}$~erg/s,
and external density profile 
$\rho(r)=10^{-25} [1+(r/35{\rm kpc})^2]^{-9/8}$~g/cm$^3$. The density contrast
$\eta$ is $6 \times 10^{-3}$ at the inlet and the
Mach number is $M=10$.

\begin{figure}[t]
\centerline{
\psfig{figure=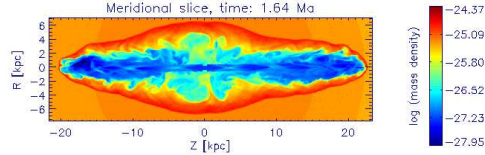,width=7cm,angle=-0}
}
\caption{Density slice for the 3D simulation. 
The jet is injected in the center in opposite 
directions.
}
\label{ml6}
\end{figure}

A slice through the computational volume at a time of 1.6~Myr is shown 
in Fig.~\ref{ml6}. 
Some basic features of such a jet are: Initially, the bow shock forms a 
spherical bubble. Then it first elongates the bubble, and after some time 
it produces cigar shaped extensions. The aspect ratio of the bow shock is 
monotonically increasing for the whole simulation. At the time shown,
one jet arm is longer than the other one by $\approx 10\%$. 
The background is slightly asymmetrically perturbed, with the higher density 
on the side of the longer jet. This shows that the asymmetry is 
caused by the action of KHIs on the jet beam.
The central part of the bow shock is pressure driven and its width turns 
out to scale as
$t^{0.6}$, for the whole simulation time. 

The observed jet in Cygnus~A has basically the same features as the simulated
one. The bow shock is a big bubble with a cigar shaped extension on one side.
The armlength asymmetry is $\approx 10\%$. But the bow-shock-to-cocoon width 
$\approx 1.3$ and the axis ratio of the bow shock ($>3$) is far-off the
observed values. 
Hence, the assumed parameters wer not a good choice. 

\section{Parameter Study}
\label{pars}
In order to find out, what parameters differ and in which direction,
we performed an axisymmetric study with constant background, 
density contrasts $\eta \in [10^{-5};10^{-2}]$, and Mach numbers of 3,~8, and 26
\citep{mypap03a}. 
\begin{figure}[t]
\centerline{
\psfig{figure=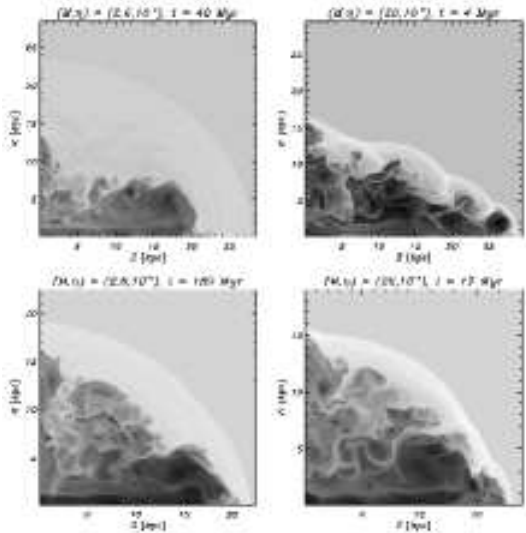,width=7cm,angle=-0}
}
\caption{Logarithmic density plots of the parameter study. Top row:
$\eta=10^{-2}$, bottom row: $\eta=10^{-4}$. Left column: M=3, right column: M=26.
Darker colour corresponds to less dense regions.
}
\label{densheet}
\end{figure}
Fig.~\ref{densheet} shows that the trend found in classical parameter studies
\citep[cocoon radius $r_\mathrm{c} \propto M \eta^{-1/4}$]{Norm93} 
is also confirmed here:
the lighter jets show fatter cocoons. However,
the cocoons of the $\eta=10^{-4}$ jets are still under-expanded, which may be the
reason for the independence of $r_\mathrm{c}$ on $M$.
The bow shocks are spherical at the beginning, because the pressure inside
the jet bubble drives them faster than the thrust at the jet head.
The relevant equation is the force balance at the bow shock, which can be transformed 
into the following integral equation
\begin{equation}\label{solu}
\int_0^r {\bf M}(r) r dr = 2 \int_0^t dt_1 \int_0^{t_1} E(t_2) dt_2
\end{equation}
that relates the mass ${\bf M}$ inside the radius $r$ to the energy $E$, injected
during the time t. For constant energy injection rate and constant density profile,
one obtains the solution of \citet{CWM75}:
$r=\left(5L/4\pi\rho_0\right)^{0.2} t^{0.6}
$. 
The critical radius for the first deviation from spherical symmetry is 
given by:
$r_1/r_\mathrm{j}=0.5 \eta^{-1/4}
$
which can be checked in the simulations.
All the simulated jets show a high cocoon pressure, which decreases in time, as expected 
for a blastwave with constant energy input.
The jet beam does not react by compression or expansion, but by adjusting its own
pressure via oblique shock waves. The result is approximate 
pressure equilibrium,
besides the hot spot.

\section{Discussion}
\label{discu}
The 3D~simulation has falsified the assumed parameters for Cygnus~A.
The observed cocoon and bow shock is wider than in the simulation. This points 
to a more extreme density contrast. It turns out from the parameter 
study that $\eta \approx 10^{-4}$ is needed in order to
get the observed values. This agrees well with the extrapolations of
\citet{Rosea99} concerning the cocoon width.
Equation \ref{solu} can be used to infer the average jet power 
($L=8\times10^{46}$~erg/s), 
and the age ($t=27$~Myr), 
which agrees with estimates in the literature.
The simulations have shown that the pressure has the same order of magnitude
in the whole system. The external pressure in the cluster gas is 
\citep{Sea01}:
$p_\mathrm{ext} \approx 3 \times 10^{-10}$~erg/cm$^3$. 
The magnetic energy density in the hot spots is 
$u_\mathrm{mag,HS}=9 \times 10^{-10}$~erg/cm$^3$
or higher
\citep{WYS00}, assuming self-synchrotron-Compton
origin of the X-ray emission. 
This is roughly the equipartition value.
Since the magnetic field in the hot spots is probably shock-amplified, this implies
a similar value of the internal energy in the jet and the cluster gas,
as predicted by the above model.
Given the density contrast of $\eta=10^{-4}$ and taking into account a possibly 
relativistic jet \cite[compare][]{Scheckea02}, we estimate:
$\Gamma^2 h \rho_\mathrm{j} c^2 \approx 10^{-8}$~erg/cm$^3$ 
($\Gamma$: bulk Lorentz factor,
$h$: specific relativistic enthalpy, $\rho_j$: mass density).
Consequently, the kinetic power of the jet is:
$L_\mathrm{kin}=2\pi r_\mathrm{j}^2 \rho_\mathrm{j} \Gamma (\Gamma h -1) \beta c^3
=5\times 10^{45} (1-1/\Gamma h)$~erg/s. ($\beta=v_\mathrm{j}/c$)
Since this falls short of the value derived above, the energy has to be in the 
magnetic power:
$L_\mathrm{B}=r_\mathrm{j}^2\Gamma^2\beta c B^2/4=5 \times 10^{44} 
(u_\mathrm{mag,j}/u_\mathrm{mag,HS})^2 \Gamma^2 $.
It follows, that $\Gamma$ has to be roughly 20.
The mass flux in the jet is then given by:
$\dot{M}=\pi r_\mathrm{j}^2 \rho_\mathrm{j} \Gamma \beta c= 3 \times 10^{-3} h^{-1} 
M_\odot$/yr.
\citet{AC88} estimate the mass flux needed in order to produce the observed 
luminosity in the Cygnus~A hotspots downstream of a MHD shock:
$\dot{M}=5\times10^{-4} \epsilon_\mathrm{rel}^{-1} M_\odot$/yr
($\epsilon_\mathrm{rel}$: energy fraction in relativistic electrons).
This does not preclude protons in the jet, and places $h$ in the regime of a few.

The 3D simulation shows that the KHI is strongest in the symmetry plane,
drawing fingers of shocked external cluster gas into the radio cocoon.
We propose that some of the X-ray enhancements inside the cocoon could be
due to this process. Outer X-ray enhancements may be caused by waves that
the KHIs excite in the shocked cluster gas.

Some cDGs show similar features like Cygnus~A. For example, the X-ray structure
of 3C~317 \citep{Blanea01} (two elliptical rings, 
elongated in the same direction) can be explained as a bubble with a
cigar shaped extension, seen at an inclination of $37^\circ$.

This work was supported by the Deutsche Forschungsgemeinschaft 
(Sonderforschungsbereich~437). 







\bibliography{c_mkrause}

\begin{thebibliography}{14}
\expandafter\ifx\csname natexlab\endcsname\relax\def\natexlab#1{#1}\fi
\expandafter\ifx\csname url\endcsname\relax
  \def\url#1{\texttt{#1}}\fi
\expandafter\ifx\csname urlprefix\endcsname\relax\def\urlprefix{URL }\fi

\bibitem[{{Appl} and {Camenzind}(1988)}]{AC88}
{Appl}, S., {Camenzind}, M., 1988,  \aap, 206, 258

\bibitem[{{Blanton} et~al.(2001){Blanton}, {Sarazin}, {McNamara}, and
  {Wise}}]{Blanea01}
{Blanton}, E.~L., {Sarazin}
	et al.,
2001,  \apjl, 558, L15

\bibitem[{{Carilli} and {Barthel}(1996)}]{CarBar96}
{Carilli}, C.~L., {Barthel}, P.~D., 1996, \aap, Review 7, 1

\bibitem[{{Castor} et~al.(1975){Castor}, {Weaver}, and {McCray}}]{CWM75}
{Castor}, J., {Weaver}, R., {McCray}, R., 1975,
  \apjl, 200, L107

\bibitem[{{Krause}(2002)}]{myphd}
{Krause}, M., 2002, PhD Thesis,
  Heidelberg, Germany,  http://www. ub.uni-heidelberg.de/archiv/2114

\bibitem[{{Krause}(2003)}]{mypap03a}
{Krause}, M., 2003, \aap, accepted, astro--ph/0211448.


\bibitem[{{Norman}(1993)}]{Norm93}
{Norman}, M.~L., 1993, in:
  Astrophysical Jets, 
  CUP, eds: Burgarella, D., Livio, M.,\& O'Dea, C.

\bibitem[{{Rosen} et~al.(1999){Rosen}, {Hughes}, {Duncan}, and
  {Hardee}}]{Rosea99}
{Rosen}, A., {Hughes}, P.~A. et al., 
1999, \apj, 516, 729

\bibitem[{{Scheck} et~al.(2002){Scheck}, {Aloy}, {Mart{\' i}}, {G{\' o}mez},
  and {M{\" u}ller}}]{Scheckea02}
{Scheck}, L., {Aloy}, M.~A. et al., 
2002, \mnras, 331, 615.

\bibitem[{{Smith} et~al.(2002){Smith}, {Wilson}, {Arnaud}, {Terashima}, and
  {Young}}]{Sea01}
{Smith}, D.~A., {Wilson}, A.~S. et al., 
2002,  \apj, 565, 195

\bibitem[{{Soker} et~al.(2002){Soker}, {Blanton}, and {Sarazin}}]{Sokea02}
{Soker}, N., {Blanton}, E.~L., {Sarazin}, C.~L., 2002,  \apj, 573, 533.

\bibitem[{{Wilson} et~al.(2000){Wilson}, {Young}, and {Shopbell}}]{WYS00}
{Wilson}, A.~S., {Young}, A.~J., {Shopbell}, P.~L., 2000,  \apjl, 544, L27.

\bibitem[{{Ziegler} and {Yorke}(1997)}]{ZY97}
{Ziegler}, U., {Yorke}, H.~W., 1997, Comp. Phys. Com. 101, 54.

\end{thebibliography}

\end{document}